\documentclass{webofc}
\usepackage[caption = false]{subfig}
\usepackage[varg]{txfonts} 
\usepackage{wrapfig}
\usepackage{comment}
\usepackage{graphicx}
\usepackage{hyperref}
\usepackage{color}

\definecolor{airforceblue}{rgb}{0.36, 0.54, 0.66}
\definecolor{dukeblue}{rgb}{0.0, 0.0, 0.61}
\definecolor{darkmidnightblue}{rgb}{0.0, 0.2, 0.4}

\hypersetup{
colorlinks=true,
    linkcolor={darkmidnightblue},
    filecolor={darkmidnightblue},      
    urlcolor={darkmidnightblue},
    citecolor = {airforceblue},
    pdftitle={airforceblue},
        }

\newcommand{\pt}{\mbox{$p_T$}\xspace}

\newcommand{\herschel }{\mbox{\textsc{HeRSCheL}}\xspace}
\newcommand{\sqsn}{\mbox{$\sqrt{s_{_{NN}}}$}\xspace}

\newcommand{\sqs}{\mbox{$\sqrt{s}$}\xspace}

\newcommand{\jpsi}{\mbox{$J/\psi$}\xspace}
\newcommand{\psip}{\mbox{$\psi(2S)$}\xspace}

\newcommand{\pnu}{Pusan National University, Busan 46241, South Korea}
\newcommand{\inha}{Inha University, Incheon 22212, South Korea}

\begin{document}
\title{Hard Probes in Ultraperipheral Collisions at LHCb}

\author{\firstname{Krista} \lastname{Smith}\inst{1,2}\fnsep\thanks{\email{krista.lizbeth.smith@cern.ch}}, on behalf of the LHCb Collaboration
}

\institute{\pnu \and \inha}

\abstract{Measurements of quarkonia production in peripheral and ultraperipheral heavy-ion collisions are sensitive to photon-photon and photon-nucleus interactions, the partonic structure of nuclei, and the mechanisms of vector-meson production. In this contribution, quarkonium measurements with the highest precision currently accessible will be compared with the latest theoretical models. Additionally, new studies of $\rho(770)$ and $\phi(1020)$ vector meson production at forward rapidity will be presented and contrasted with other experimental results. Future UPC measurements with the upgraded LHCb detector in Run 3 will also be discussed.}
\maketitle
\section{Introduction}
\label{sec:intro}
Vector meson production can be understood as an interaction between a virtual photon $\gamma^{\star}$ and two gluons (or a pomeron ${\rm I\!P}$). Vector mesons, such as the \jpsi, $\phi(1020)$, and $\rho(770)$ mesons, have the same quantum numbers as the photon ($J^{P}=1^{-}$).  
Coherent interactions, when the photon interacts with the whole nucleus, are characterized by no additional particle production. Inelastic interactions are also possible, where gluon radiation occurs, and the target or projectile nuclei dissociate from the $\gamma$-${\rm I\!P}$ interaction.  
In ultraperipheral collisions (UPC), $\gamma$-$\gamma$ and ${\rm I\!P}$-${\rm I\!P}$ interactions can also take place, which are all examples of interactions with small momentum transfer ($Q^2$). Soft processes can probe small Bj$\ddot{\mathrm{o}}$rken-$x$ values, as shown by the $Q^2$ vs. $x$ diagram in Fig.~\ref{fig:fig1}-a.  Small $x$ values imply higher gluon densities, where gluons are expected to radiate and recombine at equal rates~\cite{Aschenauer:2017jsk}. UPC could provide an environment with high levels of self-interaction and access to exotic phenomena, such as glueballs or color glass condensate, or tetraquarks.  

\section{Data Set \& Experimental Setup}
\label{sec:data}
\begin{figure}[t]
\raisebox{-1.5mm}
{\includegraphics[height=3.45cm]{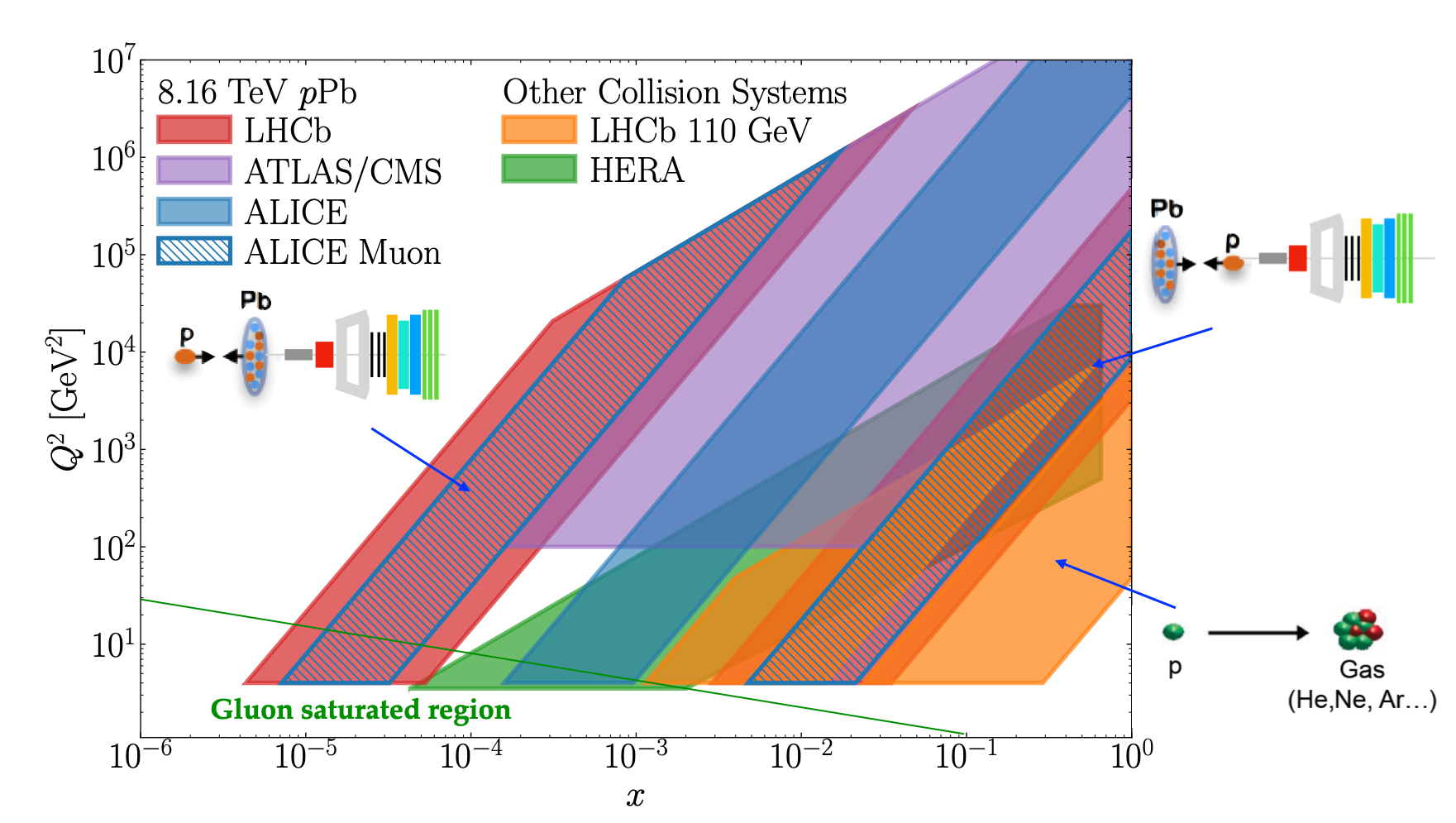}}
  \hfill
   \includegraphics[height=3.45cm]{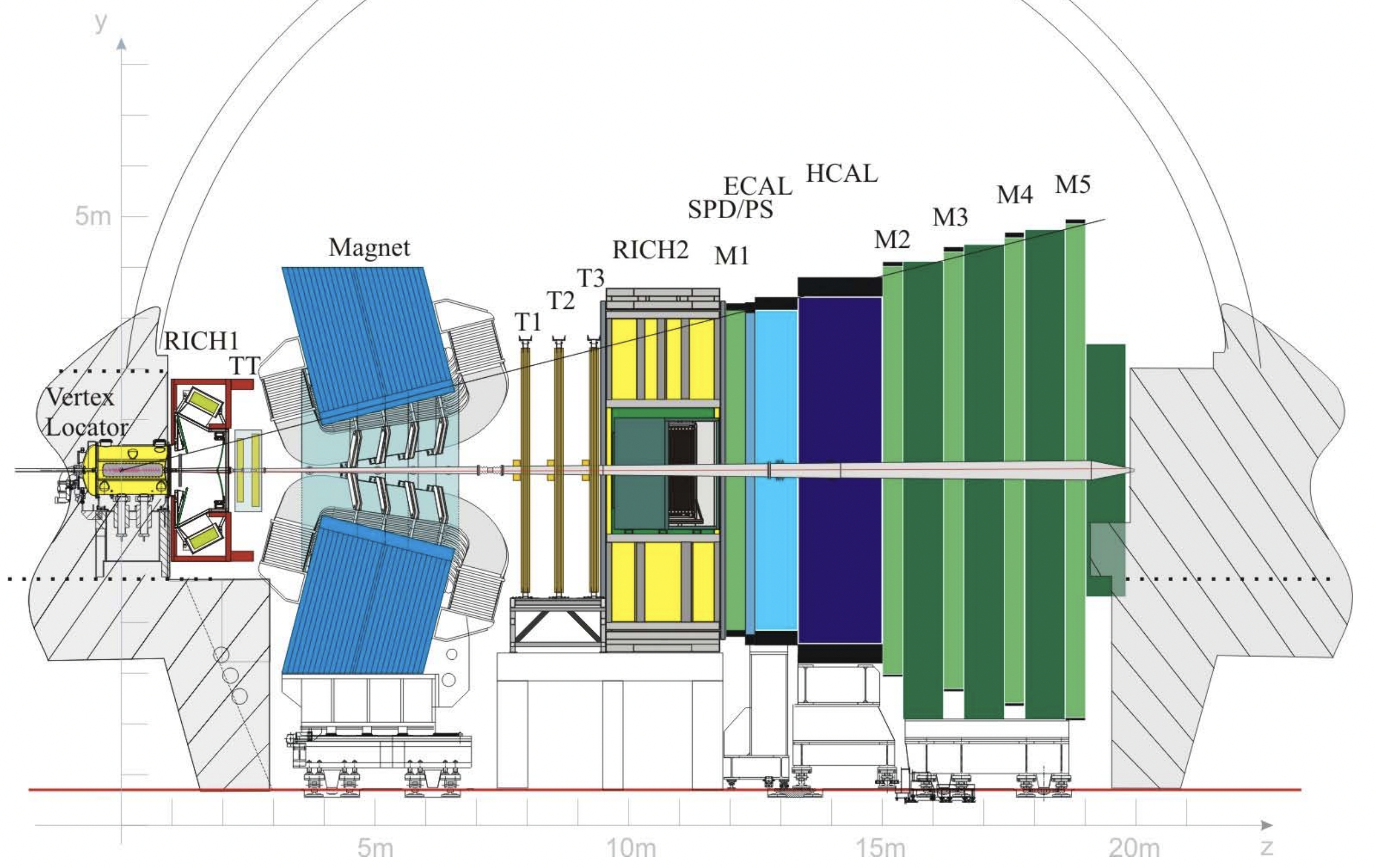}
  \caption{Left:  The $Q^2$ vs $x$ reach for HERA and the four major LHC experiments, with the LHCb collider (fixed target) mode shown by the red (orange) curve. Right:  The LHCb detector configuration for Run 2 data taking~\cite{AbellanBeteta:2020amj}.}
  \label{fig:fig1}
\end{figure}

The data samples used in the following analyses were recorded during the LHC Run 2 period (2015-2018). The PbPb data sample was recorded at a center of mass energy $\sqrt{s_{_{NN}}}$=5.02 TeV, while the $pp$ data sample was recorded at $\sqrt{s}$=13 TeV. The corresponding integrated luminosity for each sample is 228~$\mu$b$^{-1}$ and 5~fb$^{-1}$, respectively. The LHCb experiment was designed for searches of new physics in beauty and charm hadron decays. The detector covers the forward rapidity region and can measure particles from \pt~$>$ 0 GeV/c for pseudorapidity $2 < \eta < 5$. The LHCb detector configuration for Run 2 is shown in Fig.~\ref{fig:fig1}-b~\cite{AbellanBeteta:2020amj}.

\section{Centrally Produced, Peripheral, and UPC Results}
\label{sec:results}
\begin{wrapfigure}{r}{0.5\linewidth}
\includegraphics[width=\linewidth]{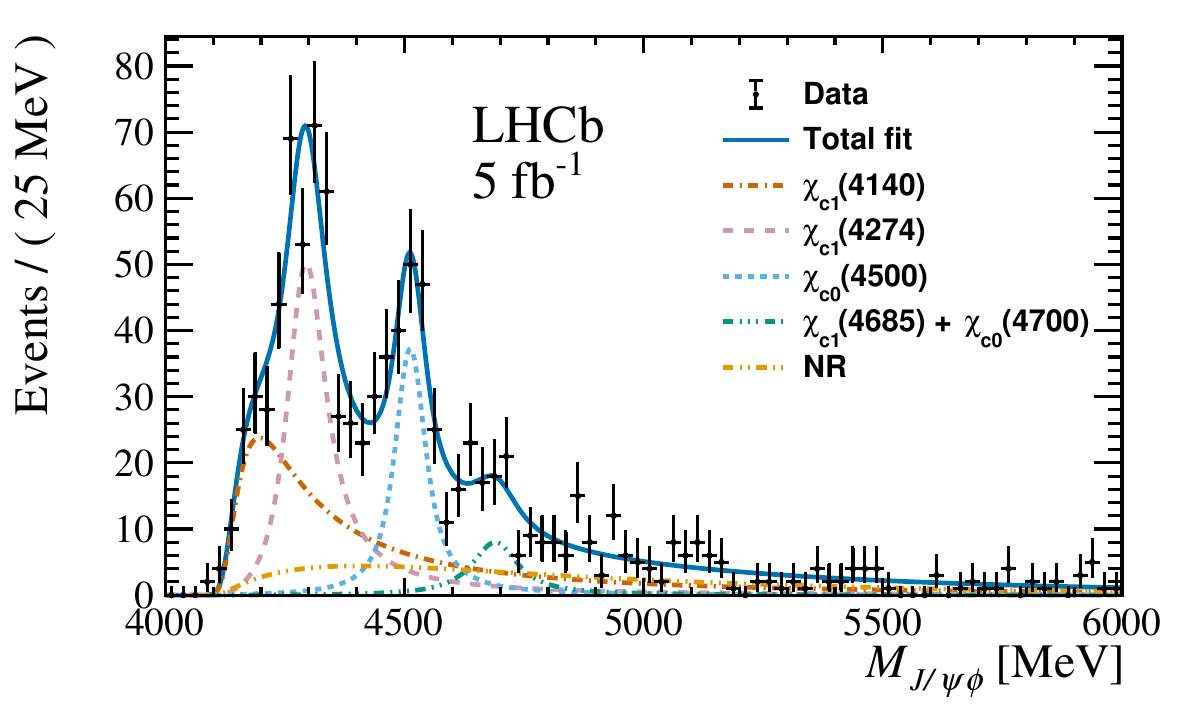}
\caption{\label{fig:fig2} Centrally produced $J/\psi\phi$ invariant mass spectrum in $pp$ collisions at \sqs$=13$~TeV~\cite{LHCb:2024smc}.}
\end{wrapfigure}

The results from hard probes in ultraperipheral collisions at the LHCb experiment are shown in Figs.~\ref{fig:fig2}-\ref{fig:fig6} and organized into the following:  proton-proton ($pp$) collisions, peripheral collisions, and ultraperipheral collision results.  

\subsection{Centrally Produced and Peripheral Collision Results}
\label{sec:}
In central diffractive collisions from $pp$ collisions at \sqs$=13$~TeV, the exotic states $\chi_{c_{0}}(4500)$ and $\chi_{c_{1}}(4274)$ were observed in the $J/\psi\phi$ invariant mass spectrum with greater than 4$\sigma$ significance~\cite{LHCb:2024smc}.  In addition to these potential tetraquark states, the signals $\chi_{c_{1}}(4140)$, 
$\chi_{c_{1}}(4685)$, and $\chi_{c_{1}}(4700)$ also appear in the spectrum, confirming previous LHCb results in hadronic collisions~\cite{LHCb:2016axx}. With the use of the \herschel detector~\cite{LHCb-DP-2016-003}, it was determined that more than two-thirds of the candidates in the spectrum are produced in events in which one or both protons dissociate.

\begin{figure}[tbh]
 \raisebox{-1.25mm}
    {\includegraphics[height=4.0cm,clip]{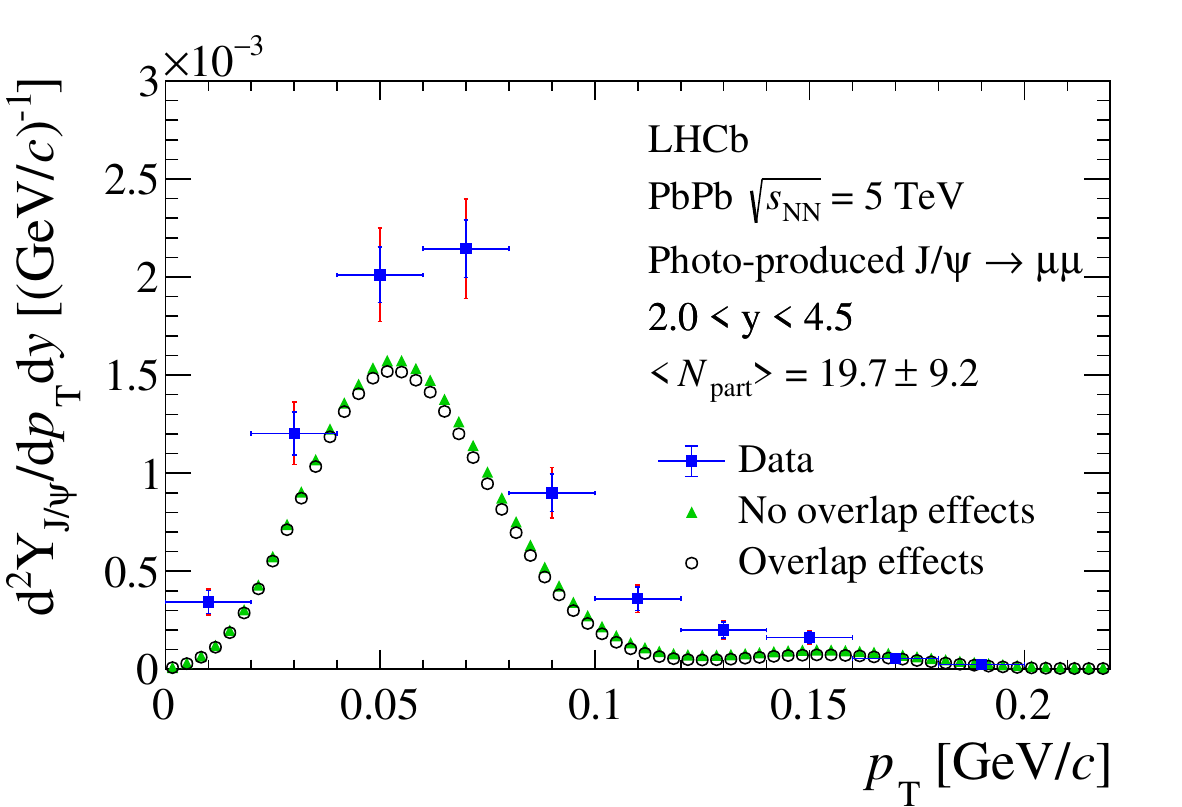}}
 \hfill
    \includegraphics[height=3.6cm]{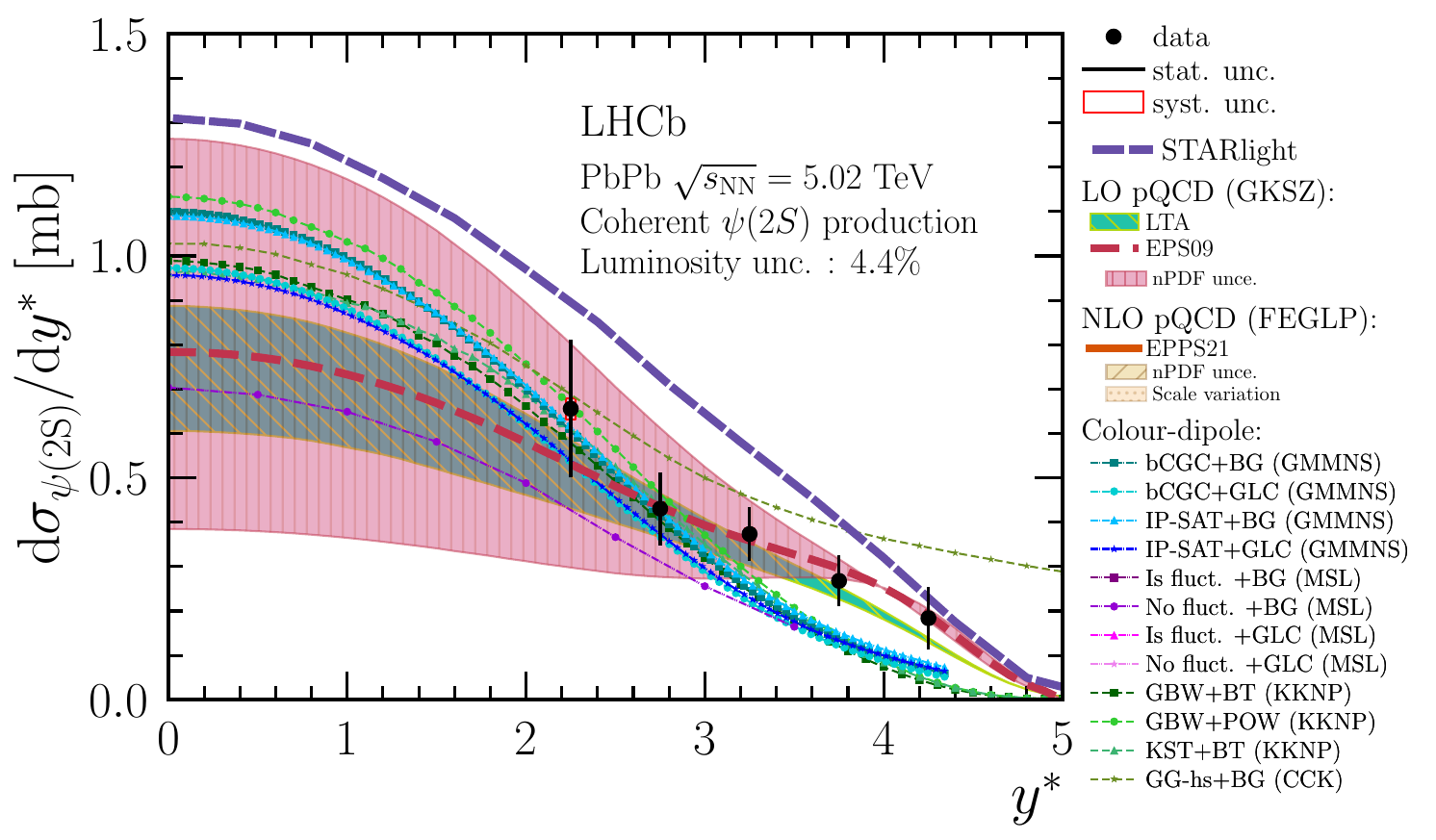}
   \caption{Left: The invariant yield as a function of $p_T$ for \jpsi photoproduction candidates~\cite{LHCb:2021hoq}.  Right: The differential cross section vs rapidity for \psip mesons in UPC at \sqsn$=5.02$~TeV~\cite{LHCb:2022ahs}. }
  \label{fig:fig3}
\end{figure}

In results from peripheral PbPb collisions, the number of \jpsi events as a function of ln$(p_T^2)$ is used to discriminate between $J/\psi$ hadronic production at the high ln$(p_T^2)$ and $J/\psi$ photoproduction in the lower region, as two distinct peaks are formed in the distribution (figure not included). In Fig~\ref{fig:fig3}-a, the invariant yield as a function of $p_T$ is shown for the \jpsi photoproduction candidates, where the mean of the distribution was found to be $\langle p_{T}\rangle=64.9\pm2.4$ MeV/c, consistent with coherent vector meson photoproduction. These results confirm earlier measurements of excess $J/\psi$ yield observed in peripheral PbPb collisions by the ALICE~\cite{ALICE:2015mzu} and STAR~\cite{STAR:2019yox} experiments.
 
\subsection{Ultraperipheral Collision Results}
\label{sec:}
\begin{wrapfigure}{l}{0.5\linewidth}
\includegraphics[width=\linewidth]{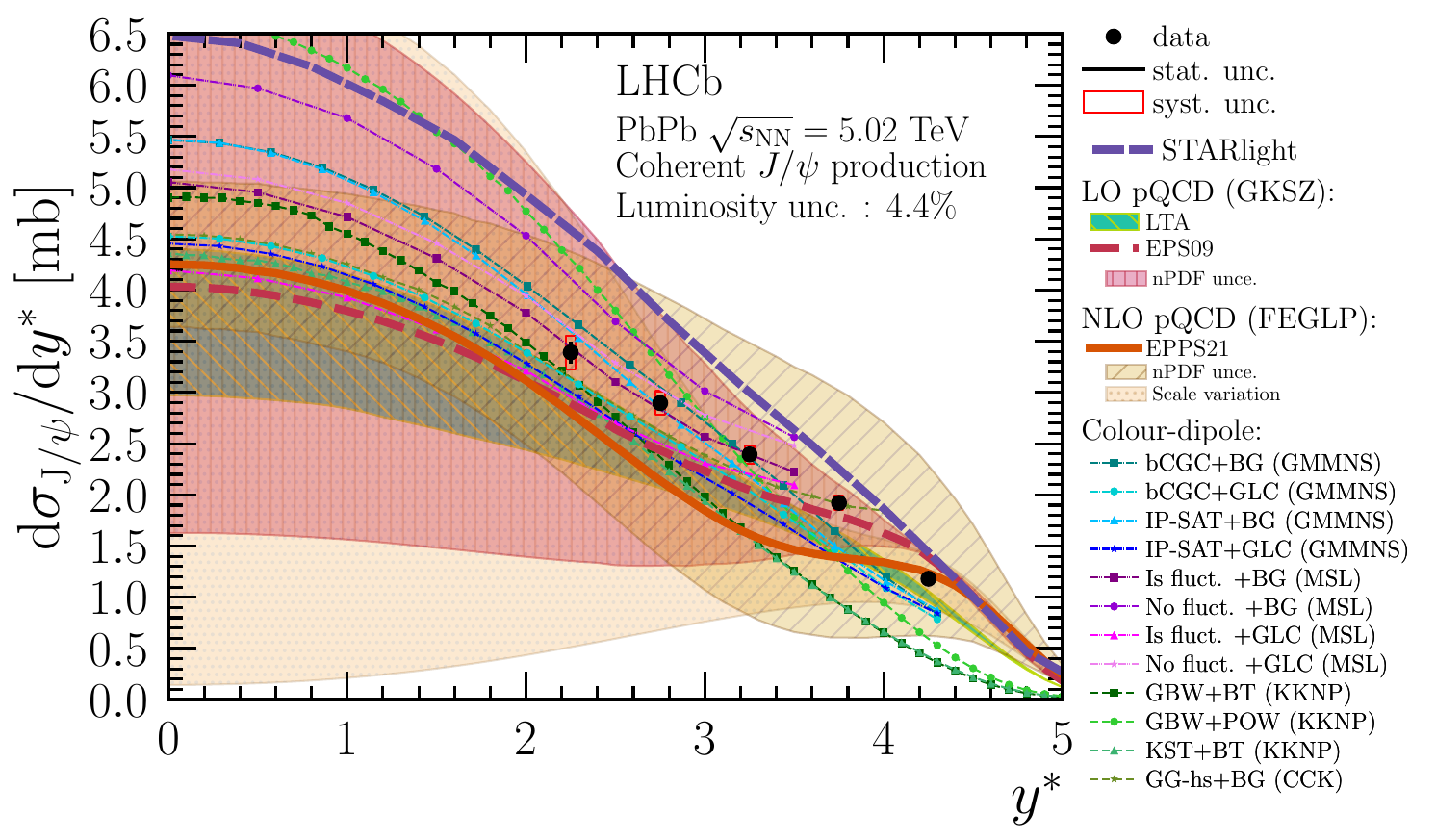}
\caption{\label{fig:fig4} The differential cross section vs rapidity for \jpsi mesons in UPC at \sqsn$=5.02$~TeV~\cite{LHCb:2022ahs}.}
\end{wrapfigure}

The first measurement of the \psip meson in UPC at forward rapidity is shown in Fig.~\ref{fig:fig3}-b. The differential cross section for \jpsi (Fig.~\ref{fig:fig4}) and \psip mesons as a function of rapidity is compared to various pQCD and color glass condensate predictions. The \psip data  
 is described reasonably well by pQCD calculations, particularly at large rapidity. 
 With its smaller statistical uncertainties, the \jpsi cross section measurement can discriminate between different predictions.

\begin{figure}[t]
\centering
\raisebox{-0.3mm}{\includegraphics[height=4.2cm]{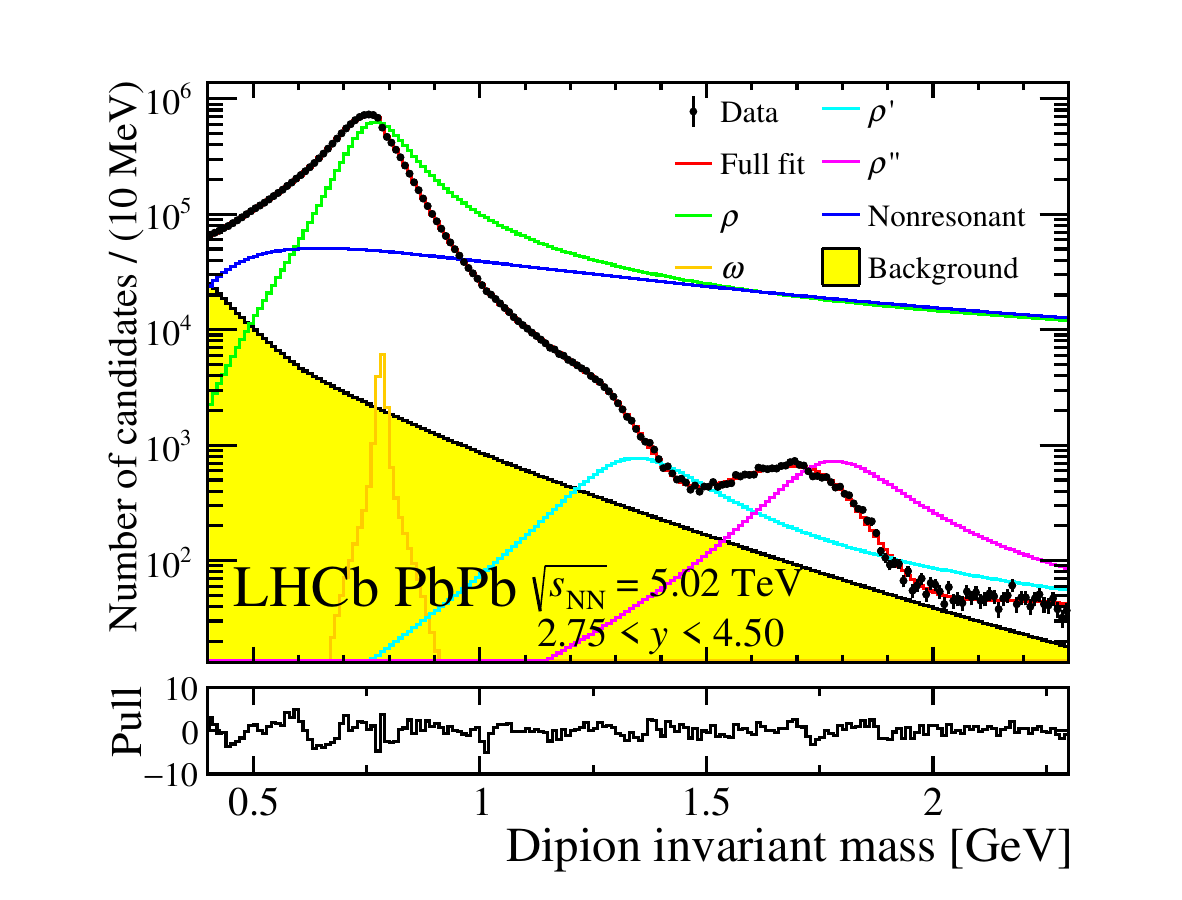}}
 \hspace{13mm}
    \includegraphics[height=3.8cm]{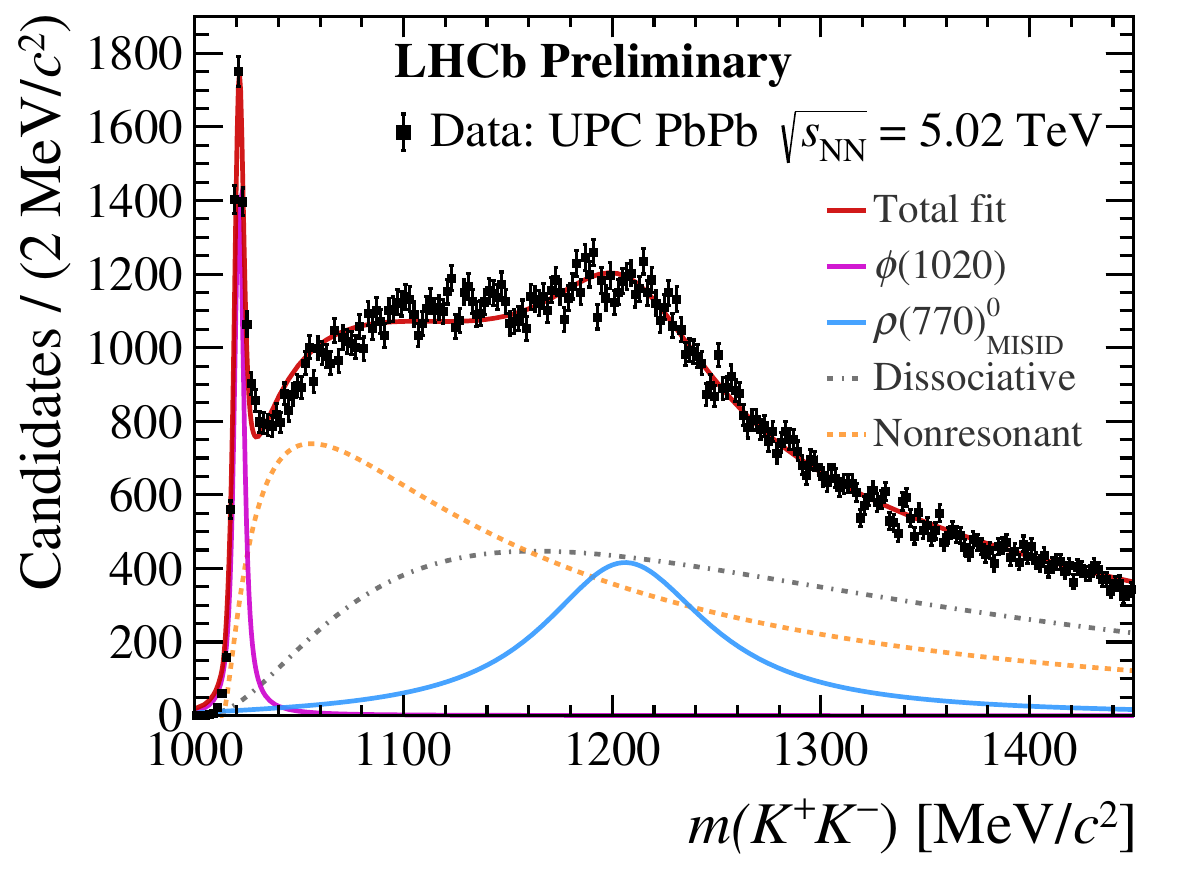}
  \caption{The full $\pi^+\pi^-$~\cite{LHCb:2025fzk} (left) and the low $K^+K^-$~\cite{LHCb:2025dim} (right) invariant mass distributions in ultraperipheral PbPb collisions at \sqsn$=5.02$~TeV for $p_T<100$~MeV/$c$.}
  \label{fig:fig5}
\end{figure}

Fig.~\ref{fig:fig5} shows the full $\pi^+\pi^-$~\cite{LHCb:2025fzk} (left) and the low $K^+K^-$~\cite{LHCb:2025dim} (right) invariant mass distributions in ultraperipheral PbPb collisions at \sqsn=5.02~TeV for $p_T<100$~MeV/$c$. The $\pi^+\pi^-$ distribution corresponds to approximately 20 million candidates and can be compared to previous measurements at midrapidity by ALICE~\cite{ALICE:2020ugp}.  Both experiments observe an excess of events near 1.7~GeV/$c^2$, where the LHCb collaboration determined the structure is consistent with the $\rho(1450)$ and $\rho(1700)$ mesons. In the $K^+K^-$ distribution, the first observation of the $\phi(1020)$ meson in UPC at forward rapidity is shown, with significance greater than 5$\sigma$ (see also~\cite{CMS:2025lsm}). The broader structure at higher invariant mass is interpreted as misidentified pions from $\rho(770)^0 \rightarrow \pi^+\pi^-$ decays.

\begin{wrapfigure}{r}{0.5\linewidth}
\includegraphics[width=\linewidth]{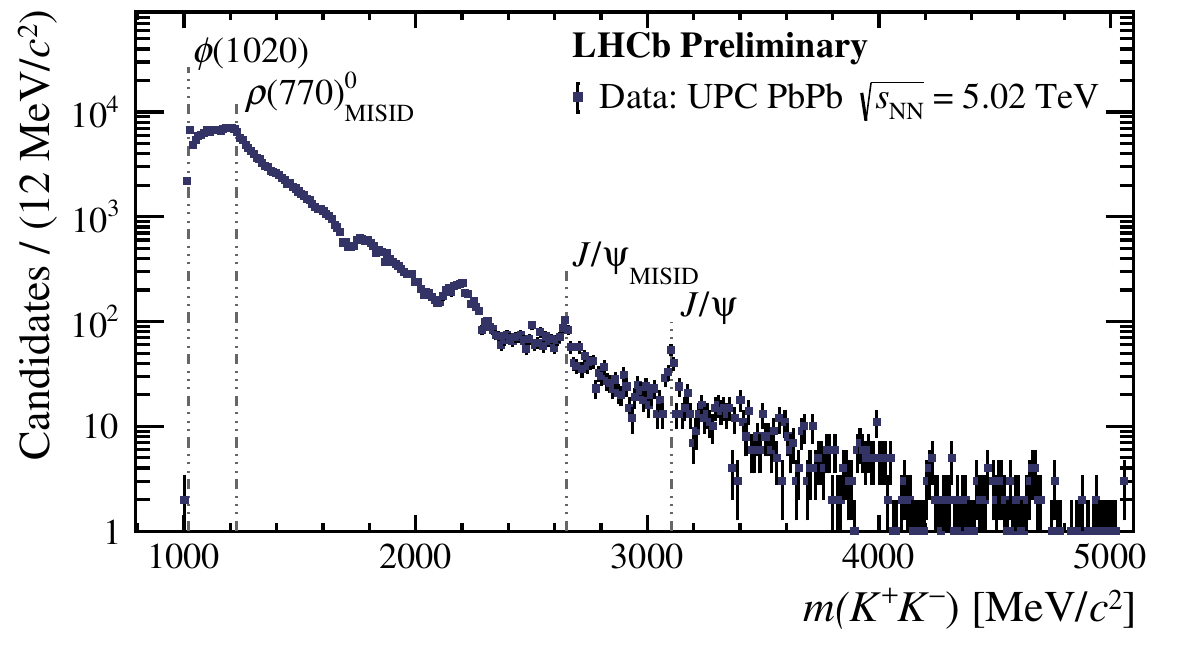}
\caption{\label{fig:fig6} The full $K^+K^-$ invariant mass distribution in UPC at \sqsn$=5.02$~TeV~\cite{LHCb:2025dim}.  The $\phi(1020)\rightarrow K^{+}K^{-}$ signal at the lower end of the spectrum was observed with $>5\sigma$ significance. }
\end{wrapfigure}

In Fig.~\ref{fig:fig6}, the full $K^+K^-$ spectrum is shown, where several peaking structures are seen at higher invariant mass, including \jpsi $\rightarrow K^+K^-$ signal and misidentified protons from \jpsi $\rightarrow p\bar{p}$. The $K^+K^-$ distribution differs from ALICE midrapidity results~\cite{ALICE:2023kgv}, showing a rich spectrum in the forward region and a more detailed study is needed to quantify these results.

\section{Conclusion}
\label{sec:conclusion}
In conclusion, the LHCb collaboration has presented results in centrally produced, peripheral, and ultraperipheral collisions. In $pp$ collisions, the $\chi_{c_{0}}(4500)$ and $\chi_{c_{1}}(4272)$ states were observed with $>4\sigma$ significance in diffractive processes estimated to contain $\sim$69\% inelastic production.  Possible future studies for Run 3 include searches for resonances in PbPb collisions through the $\gamma\gamma$ production channel. LHCb has also observed the $\phi(1020)\rightarrow K^{+}K^{-}$ signal in UPC events at \sqsn$=5.02$ TeV with $>5\sigma$ significance.  The $\pi^{+}\pi^{-}$ invariant mass spectrum appears similar at both mid and forward rapidities, and an excess of events is seen near $\sim$1.7~GeV/$c^{2}$ consistent with the $\rho(1450)$ and $\rho(1700)$ mesons. However, the $K^{+}K^{-}$ spectrum 
differs between mid and forward rapidity, with a rich spectrum seen only in the forward region. During Long Shutdown II, LHCb tracking was fully upgraded with a new Vertex Locator, Upstream Tracker (UT), and Scintillating Fiber (SciFi) detectors. A new fixed target System for Measuring Overlap with Gas (SMOG2) was also installed, in addition to other upgrades for Run 3 (2022-2025) data taking~\cite{LHCb:2023hlw}. Future LHCb Run 3 studies via the high luminosity fixed target mode with SMOG2 are possible, as well as opportunities to explore photoproduction with the upgraded photon trigger.

\bibliography{main.bib}

\end{document}